\documentclass[10pt,onecolumn,twoside]{article}

\usepackage{epsfig}
\usepackage{titlesec}
\usepackage{url} 
\usepackage{booktabs} 
\usepackage{amsfonts} 
\usepackage{amssymb} 
\usepackage{nicefrac} 
\usepackage{microtype} 
\usepackage{mathrsfs}
\usepackage{bm} 
\usepackage{cite} 
\usepackage{comment}
\usepackage{diagbox}
\usepackage{graphicx} 
\usepackage{mathtools} 
\usepackage{algpseudocode}
\usepackage{algorithm}
\usepackage{amsthm} 
\usepackage{enumitem} 
\usepackage{multirow} 
\usepackage{tabularx} 
\usepackage{hhline}
\usepackage{arydshln} 
\usepackage{enumitem} 
\usepackage{siunitx}
\usepackage[font=small,skip=5pt]{caption}
\usepackage{parskip}
\usepackage{comment}
\usepackage[dvipsnames]{xcolor}
\usepackage{fullpage}
\usepackage[hidelinks]{hyperref}

\algtext*{EndIf}
\algtext*{EndFor}
\algtext*{EndWhile}

\newcolumntype{L}[1]{>{\raggedright\arraybackslash}p{#1}}
\newcolumntype{C}[1]{>{\centering\arraybackslash}p{#1}}
\newcolumntype{R}[1]{>{\raggedleft\arraybackslash}p{#1}}

\theoremstyle{plain} 

\def\lim{\mathop{\mathsf{lim}}} 

\def\gbm{{\bm{g}}}

\def\nbm{{\bm{n}}}
\def\ubm{{\bm{u}}}

\def\gbm{{\bm{g}}}
\def\ubm{{\bm{u}}}
\def\nbm{{\bm{n}}}
\def\Rcbm{{\bm{\mathcal{R}}}}
\def\Acbm{{\bm{\mathcal{A}}}}
\def\Hcbm{{\bm{\mathcal{H}}}}

\usepackage[hidelinks]{hyperref}

\usepackage{upgreek}

\usepackage{wrapfig}

\definecolor{pink}{HTML}{000000}
\definecolor{lightgreen}{RGB}{229,251,229}
\definecolor{red}{rgb}{0,0,0}

\title{Dual-Cycle: Self-Supervised Dual-View Fluorescence Microscopy\\Image Reconstruction using CycleGAN}

\author{Tomas~Kerepecky$^{\dagger}$
\thanks{The Czech Academy of Sciences, Institute of Information Theory and Automation, Prague, Czechia}
\hspace{0.05em},
Jiaming~Liu%
\thanks{Computational Imaging Group (CIG), Washington University in St. Louis, St. Louis, MO 63130}
\hspace{0.05em}, 
Xue Wen Ng%
\thanks{Department of Cell Biology and Physiology, Washington University School of Medicine, St.\ Louis, MO 63110},
\hspace{0.05em} \\
David~W.~Piston$^{\ddagger}$,
and Ulugbek~S.~Kamilov$^{\dagger}$}

\begin{document}
\date{}
\maketitle

\begin{abstract}
Three-dimensional fluorescence microscopy often suffers from anisotropy, where the resolution along the axial direction is lower than that within the lateral imaging plane. We address this issue by presenting Dual-Cycle, a new framework for joint deconvolution and fusion of dual-view fluorescence images. Inspired by the recent Neuroclear method, Dual-Cycle is designed as a cycle-consistent generative network trained in a self-supervised fashion by combining a dual-view generator and prior-guided degradation model. We validate Dual-Cycle on both synthetic and real data showing its state-of-the-art performance without any external training data.
\end{abstract}

\section{Introduction}
\begin{wrapfigure}{r}{0.47\textwidth}
\centering
\begingroup
\setlength{\tabcolsep}{1pt}
\begin{tabular}{cc}
\includegraphics[width=0.46\linewidth]{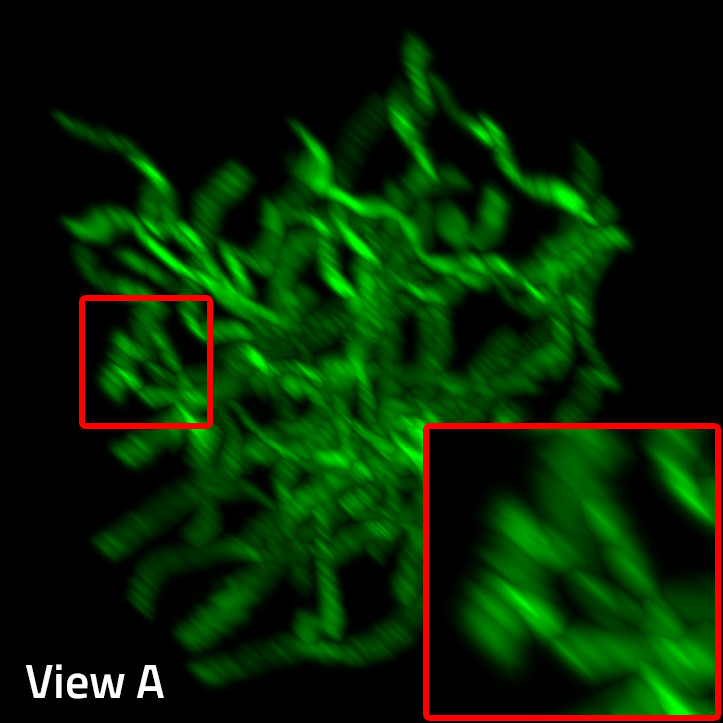}%
& \includegraphics[width=0.46\linewidth]{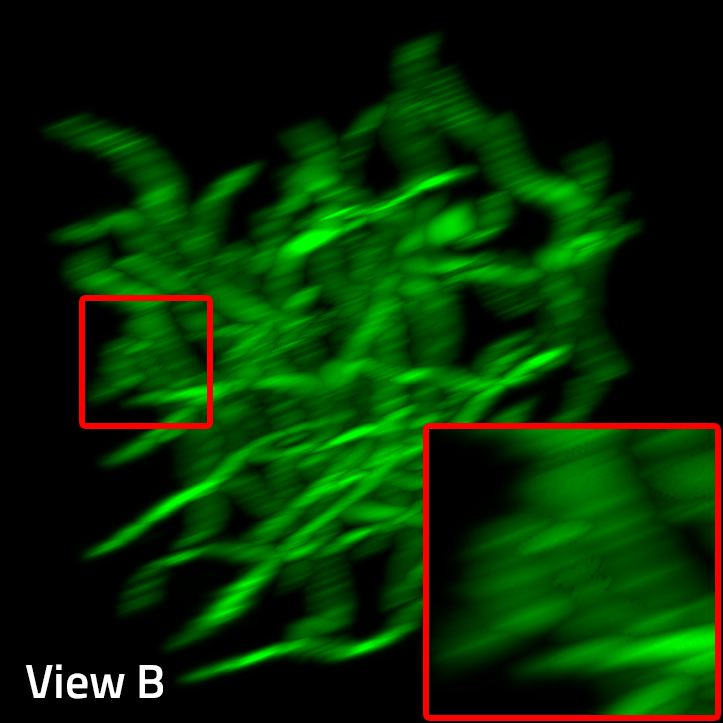}\\
\includegraphics[width=0.46\linewidth]{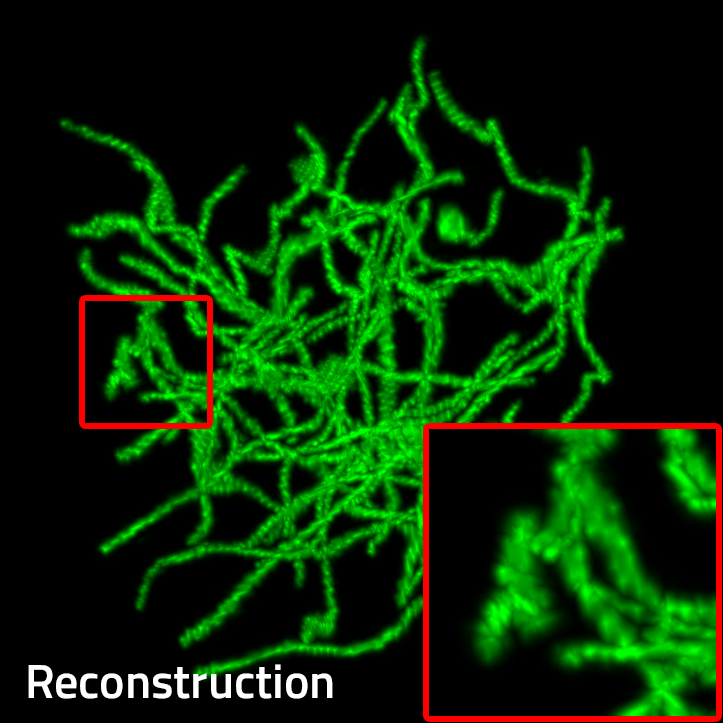}%
& \includegraphics[width=0.46\linewidth]{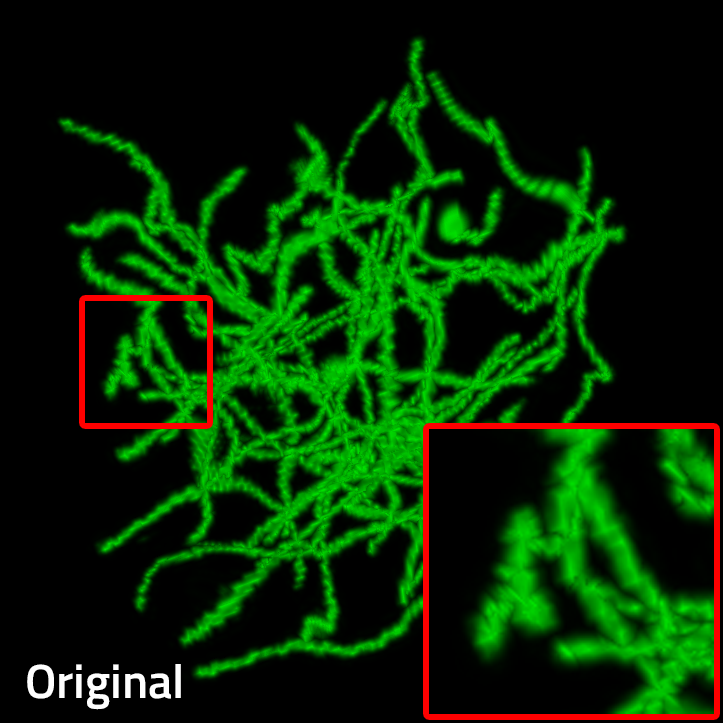}
\end{tabular}
\endgroup
\caption{Dual-Cycle reconstructs a 3D image with isotropic resolution given two views, A and B, of the same sample. \label{fig:header}}

\end{wrapfigure}

\label{sec:intro}
Three-dimensional fluorescence imaging, such as light-sheet fluorescence microscopy (LSFM)~\cite{stelzer2021light,liu2022recovery} is an essential tool for revealing important structural information in biological samples. However, it is common for 3D fluorescence microscopy to suffer from spatial-resolution anisotropy, where the axial direction is more blurry than the lateral imaging plane. Such anisotropy is due to several factors, including the diffraction of light and axial undersampling.

The spatial-resolution anisotropy is often addressed using image deconvolution methods, such as Richardson-Lucy algorithm~\cite{richardson1972bayesian,lucy1974iterative}. However, achieving isotropic resolution from a single 3D volume is an ill-posed inverse problem. The problem can be simplified by using multiview microscopy systems, such as dual-view inverted selective plane illumination microscope (diSPIM)~\cite{wu2013spatially,kumar2014dual}, equipped with classical joint multi-view deconvolution and fusion methods~\cite{wu2013spatially,preibisch2014efficient, temerinac2011multiview}.

Deep learning (DL) has emerged as an alternative to the classical deconvolution algorithms~\cite{guo2020rapid,park2022deep,wu2021multiview}. Neuroclear~\cite{park2022deep} is a recent self-supervised DL framework that uses cycle-consistent generative adversarial network (CycleGAN)~\cite{zhu2017unpaired} to improve the axial resolution from a single 3D input image without any knowledge of the point spread function (PSF). However, in many cases, the experimental PSF can be readily measured using  either fluorescent  beads~\cite{huygens, temerinac2011multiview} or small structures within samples~\cite{de2003image}, or derived theoretically~\cite{becker2019deconvolution}.

In this paper, we present Dual-Cycle as an improvement to Neuroclear that extends it into a dual-view self-supervised model-based framework. The inclusion of an additional view as input improves the reconstruction capability, while the  additional prior on estimated PSFs allows our model to account for the expected degradation process. We experimentally validate Dual-Cycle on synthetic and real data showing that it can outperform Neuroclear as well as traditional dual view reconstruction algorithms.

\begin{figure*}[!t]
  \includegraphics[width=\textwidth]{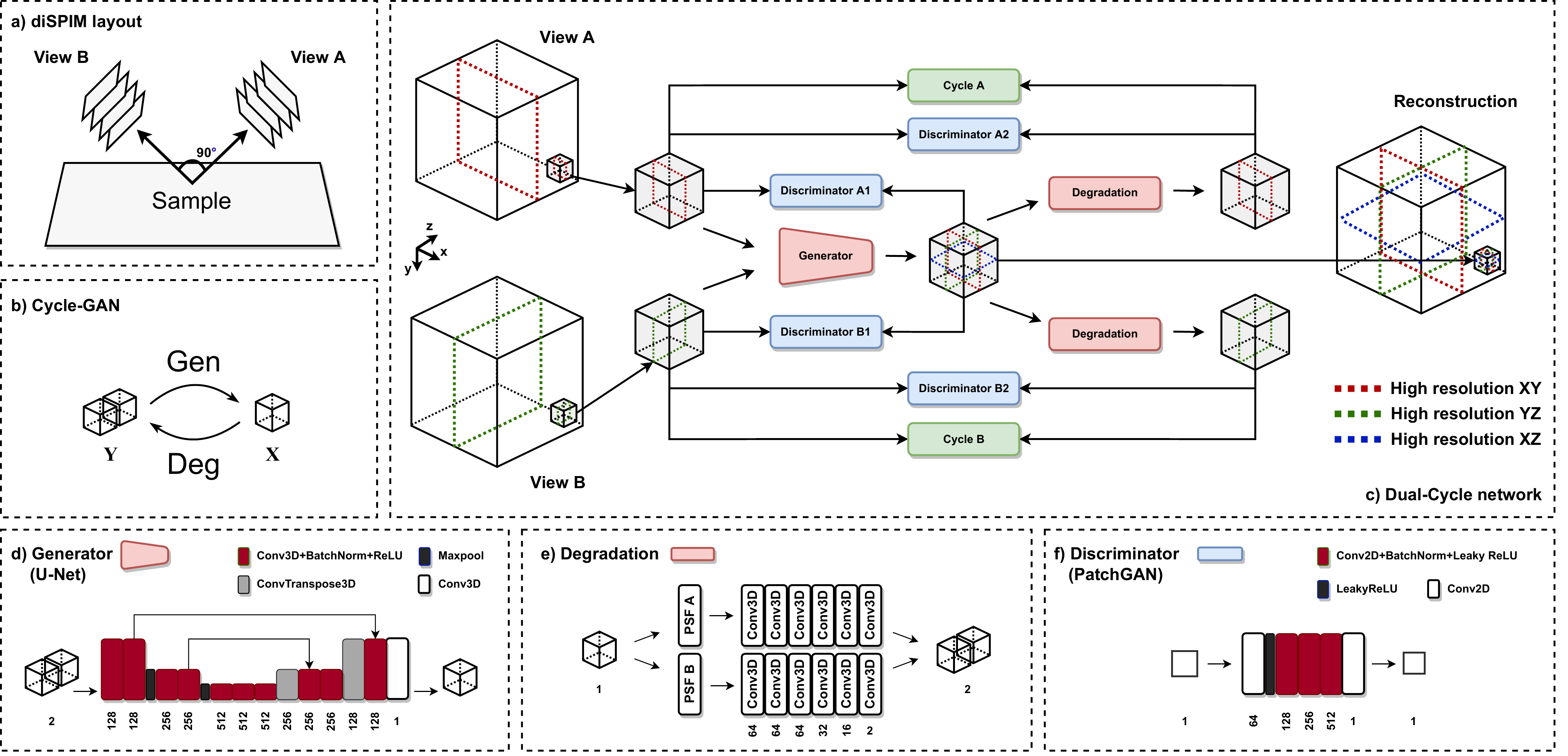}
  \caption{Schematic illustration of the Dual-Cycle framework. a) Scheme of dual-view inverted selective plane illumination microscope (diSPIM). b)~CycleGAN approach: for two domains $Y$ and $X$, CycleGAN learns two mutually-inverse generator mappings $Gen$ and $Deg$ with the assistance of corresponding discriminators. c)~Dual-Cycle network architecture. d)~Schematic of the generator based on U-Net. e)~Degradation forms two paths each consisting of blurring with known PSF followed by the deep linear generator. f)~PatchGAN-based \cite{isola2017image} discriminators work on 2D slices of input 3D volumes.} \label{fig:framework}
\end{figure*}

\section{Forward Problem}
\label{sec:backg}
We focus on images recorded with single-plane illumination
microscopes (SPIMs) \cite{huisken2004optical} in a dual-view setup (diSPIM, Fig.~\ref{fig:framework}a). Data is acquired by two cameras,  A and B, with an ideal relative rotation of 90~degrees.  The image formation process (forward model) can be represented as the following linear observation model:
\begin{align}
\begin{split}
\gbm_A &= \Acbm_A \Hcbm_A \ubm + \nbm,\\
\gbm_B &= \Rcbm_{\perp} \Acbm_B  \Hcbm_B  \ubm + \nbm.
\end{split}
\label{eq:model2}
\end{align}
where $\gbm_A$, $\gbm_B$, and $\ubm$ correspond to the vectorized forms of deskewed 3D volumes, measured by camera A (View~A), camera B (View~B), and the original high-resolution 3D volume (Fig.~\ref{fig:header}). $\Hcbm_A$ (resp.\ $\Hcbm_B$) denote 3D convolution along the axial direction $z$ (resp.\ $x$) with some known PSF $h_A$ (resp.\ $h_B$). To model the mismatch from an ideal dual view setup, we include operators $\Acbm_{A/B}$, representing 3D affine transformation. We assume a coordinate system of unknown image $\ubm$ to be the same as $\gbm_A$ and that the ideal rotation of View B  with respect to View A is 90 degrees around axes $y$, denoted as $\Rcbm_{\perp}$. We omit subsampling in the axial directions by interpolating measurements to have voxels of equal size. In the general case, we consider additive noise $\nbm$.

Problem \ref{eq:model2} leads to an inherently ill-posed inverse problem. To solve it, we adopt and extend the approach proposed in \cite{park2022deep}.

\section{Inverse Problem}
\label{sec:propo}

Our proposed framework is illustrated in  Fig.~\ref{fig:framework}c. In our setup, View A has a higher resolution in the $xy$ plane and is blurred in the axial direction $z$, while View B has a higher resolution in the $yz$ plane and is blurred in the axial direction $x$. Our goal is to reconstruct the original 3D volume with an isotropic resolution. We focus mainly on joint deconvolution and fusion with additional fine registration. 
Our framework is based on a CycleGAN approach illustrated in Fig.~\ref{fig:framework}b and consists of two cycle-consistency paths, hence the name Dual-Cycle. It is worth mentioning that Dual-Cycle does not require any external training data beside the test object to be reconstructed. 

The two views of the 3D volume are used as input for the 3D U-net-based generator (Fig.~\ref{fig:framework}d). The result of the generator is one 3D image representing the original 3D volume with isotropic resolution. To achieve this, we employ two sets of discriminators A1 and B1 (Fig.~\ref{fig:framework}f). Discriminators A1 distinguish between $xy$ planes of View A and $xy$ and $xz$ planes of the reconstructed volume.
Discriminators B1 distinguish between $yz$ planes of View B and $yz$ and $xz$ planes of the reconstructed volume. 
To regularize and stabilize learning, the dual-cycle consistency is imposed. Therefore, the reconstructed image is degraded along two paths to imitate the forward problem \eqref{eq:model2}. 
Consequently, \textit{Degradation} A and B, Fig.~\ref{fig:framework}e, consist of 3D convolution with given PSFs $h_A$ and $h_B$ followed by a deep linear generator (DLG) to address ideal model mismatch caused by affine operators $\Acbm$. For the blind case, when PSFs are unknown, degradation can be performed by DLGs only. Eventually, two other sets of discriminators A2 and B2 are added to map the distribution of corresponding planes of input View A/B onto generated View A/B. All discriminators are PatchGAN-based \cite{isola2017image} and work on 2D slices of analyzed 3D volumes (Fig.~\ref{fig:framework}f). Pixel-wise L1 loss between View A/B and generated View A/B is added to the GAN objective function to enforce cycle consistency. 

\section{Experimental Validation}
\label{sec:exper}

We now present the numerical evaluation of Dual-Cycle on synthetic and real light-sheet data.

\begin{figure}[t!]
\centering
\begingroup
\setlength{\tabcolsep}{1pt}
\begin{tabular}{c}
\includegraphics[width=0.5\linewidth]{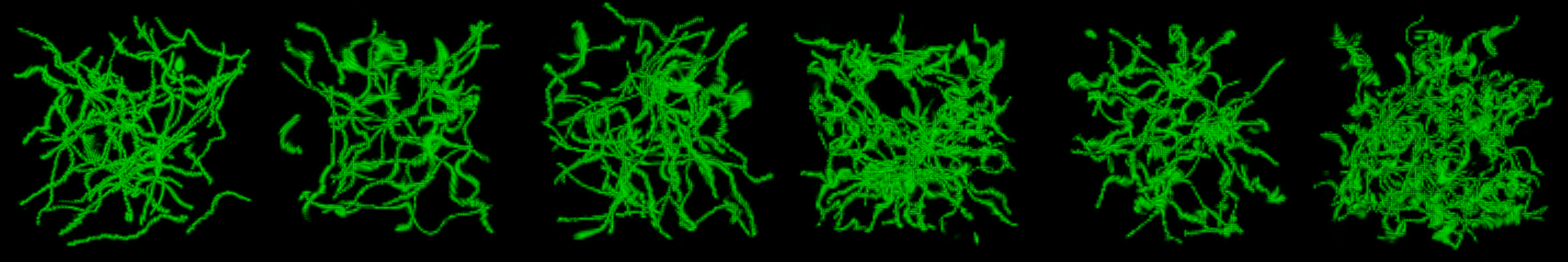}%
\end{tabular}
\endgroup
\caption{The set of six generated 3D volumes used in experiments. \label{fig:synSet}}
\centering
\begingroup
\setlength{\tabcolsep}{1pt}
\begin{tabular}{c}
\includegraphics[width=0.5\linewidth]{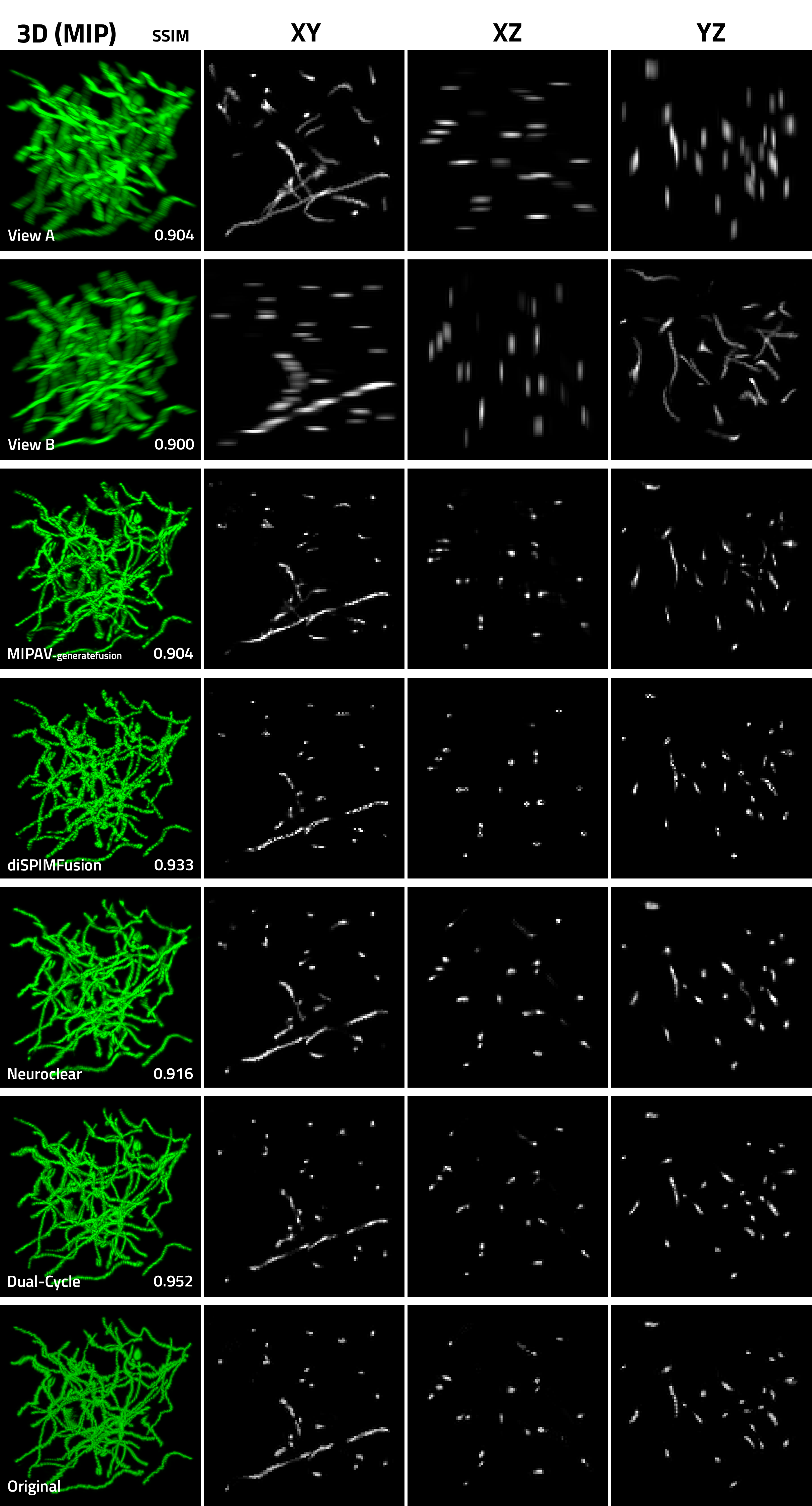}%
\end{tabular}
\endgroup
\caption{Comparison of MIPAV-generatefusion \cite{kumar2014dual}, diSPIMFusion \cite{guo2020rapid}, Neuroclear \cite{park2022deep}, and Dual-Cycle applied on the views A and B generated from the first 3D volume in the synthetic dataset in Fig.\ \ref{fig:synSet}. Visualized XY, XZ, and YZ images represent central cross-sections of the corresponding cubes in $xy$, $xz$, and $yz$ planes. Each reconstruction is labeled with its SSIM value with respect to the original volume. \label{fig:synresult}}
\end{figure}

\subsection{Synthetic data}
\label{ssec:syn}
We first illustrate possible improvements due to our dual-view framework over the single-view Neuroclear~\cite{park2022deep}. Additionally, we compare our network with other commonly used multi-view reconstruction techniques diSPIMFusion~\cite{guo2020rapid} and MIPAV-generatefusion~\cite{kumar2014dual}. The performance was measured using the peak signal-to-noise ratio (PSNR) and structural similarity index measure (SSIM).

We consider a dataset of six generated 3D volumes (120 × 120 × 120 voxels), shown in Fig.\ \ref{fig:synSet}. We drew 30-50 lines randomly in space and applied 3D elastic grid-based deformation. These volumes were treated as original ground truth volumes. All images were scaled to have values in the range 0-1. To obtain degraded volumes View A/B, we used the degradation process \eqref{eq:model2}, without noise and 90-degree rotation. The original volume was blurred in the $z$ direction for View A and in the $x$ direction for View B (blurring by Gaussian kernel with a standard deviation in range 2-4). Further, we applied random affine transformations to simulate the imperfection of the registration method. Relatively small mismatch (representing by $\Acbm$ in eq. \eqref{eq:model2}) is implemented as transformation of 3D points $\bm{p}$ as follows: $\bm{p}^{\prime} = (\mathbb{I} + \mathbb{N})\bm{p} + \bm{t}$, where $\mathbb{I}$ is identity matrix and $\mathbb{N}$ is random matrix with elements from a uniform distribution over $[-0.0025, 0.0025]$, and $\bm{t}$ is random translation vector sampled from a uniform distribution over $[-0.05, 0.05]^3$.

Except for Neuroclear, all methods use prior knowledge about the PSFs and both views as input. Visual comparison of reconstructed volumes corresponding to the first 3D volume of the synthetic dataset is in Fig.\ \ref{fig:synresult}. All methods can effectively perform the reconstruction, yet the improvement of Dual-Cycle compared to single view baseline is visually noticeable and corroborated by an increase in SSIM. Table \ref{table:results} summarizes the average PSNR/SSIM results of the tested methods. Overall, Dual-Cycle improves over the second best methods by 1.49 db (PSNR) and 0.017 (SSIM).

Implementation of Dual-Cycle was based on the Neuroclear and CycleGAN PyTorch framework; we used Adam optimizer and learning rate set to 0.0001. The network was initialized with weights pre-trained on the first volume. The training of the first (resp.\ following volumes) lasted approximately 12 hours (resp.\ 3-6 hours) using NVIDIA RTX A5000.

\subsection{Real data}
\label{ssec:subhead}

\begin{figure*}[!ht]
  \includegraphics[width=\textwidth]{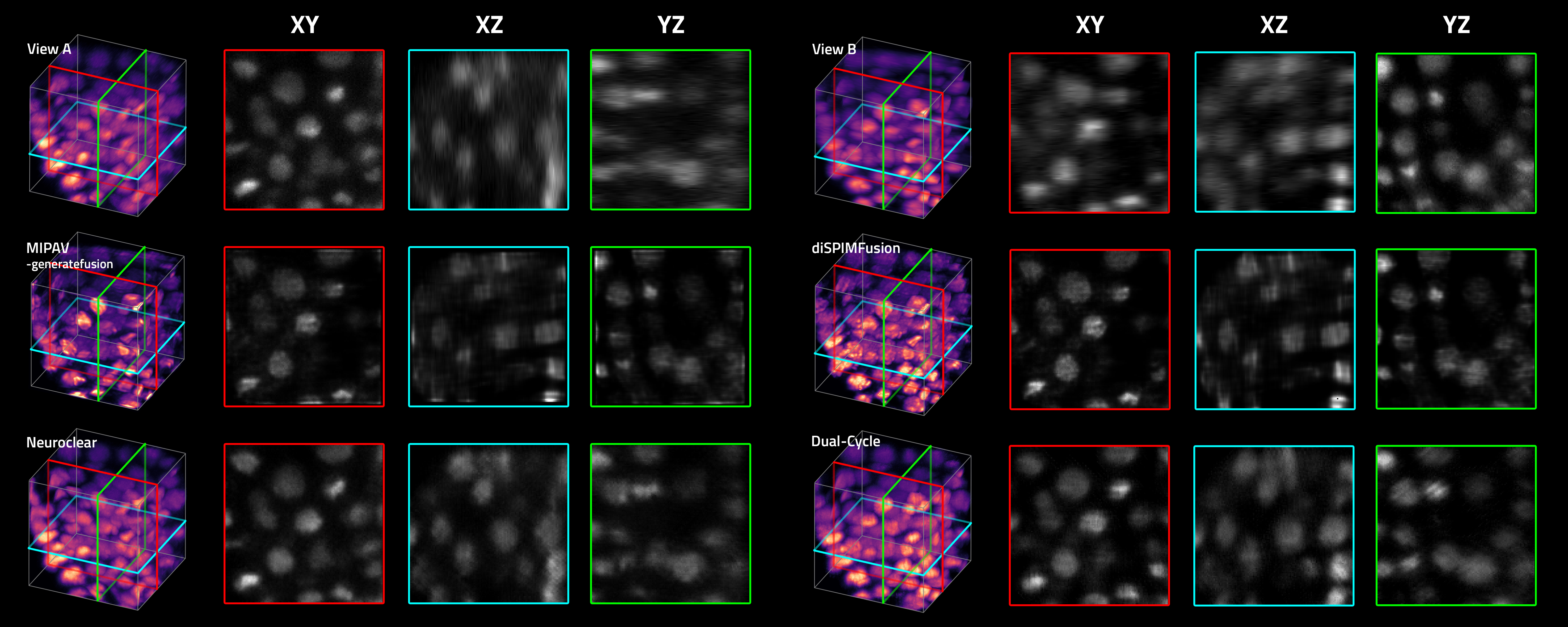}
  \caption{Image reconstruction from real diSPIM data from \cite{guo2020rapid} using  reconstruction methods MIPAV-generatefusion \cite{kumar2014dual}, diSPIMFusion \cite{guo2020rapid}, Neuroclear \cite{park2022deep}, and the proposed Dual-Cycle framework.  \label{fig:realresult}}
\end{figure*}

We also tested reconstruction on diSPIM data from \cite{guo2020rapid}. Data was preprocessed using the Fiji software \cite{schindelin2012fiji}. The preprocessing involved denoising of both views and performing initial coarse registration of View B on View A. For both views: the minimum brightness value was truncated at value 78, volumes were normalized to 0-1 range, and were interpolated to have voxel sizes equal to $(0.1625~\mathrm{\upmu m})^3$. For the registration of view B on view A, we used Fiji plugin Fijiyama \cite{fernandez2021fijiyama}.
Images were cropped to 120 × 120 × 120 voxels and tested with the same methods as in Sec.\ \ref{ssec:syn}.
Visual comparison of reconstructed volumes is presented in Fig.\ \ref{fig:realresult}. The improvement of Dual-Cycle reconstruction over the Neuroclear is indicated via cross sections. Overall, Dual-Cycle achieves comparable or better performance relative to the state-of-the-art methods.

\begin{table}[t!]
  \caption{The average PSNR/SSIM results of the blurred view~A/B, MIPAV-generatefusion, diSPIMFusion, Neuroclear and Dual-Cycle on the testing 3D volumes.\vspace{5px}}
  	\centering
  	\resizebox{0.5\columnwidth}{!}{\begin{tabular}{lrr}
	\hline\hline
 	\multicolumn{1}{c}{Method} & \multicolumn{1}{c}{PSNR [dB]} & \multicolumn{1}{c}{SSIM}\\
 	\hline
View A                                                  & 29.32  & 0.929 \\
View B                                                  & 29.13  & 0.927 \\
MIPAV-generatefusion \cite{kumar2014dual}               & 29.13  & 0.931 \\
diSPIMFusion \cite{guo2020rapid}                        & 28.55  & 0.943 \\
Neuroclear \cite{park2022deep}                          & 29.79  & 0.942  \\
Dual-Cycle (our)                                        & \textbf{31.28}  & \textbf{0.960}\\
	[1ex] \hline
	\end{tabular}}
	\label{table:results}
\end{table}

\newpage

\section{Conclusion}
We presented Dual-Cycle, a self-supervised framework for dual-view fluorescence image reconstruction. The proposed method extends the recent Neuroclear method based on the CycleGAN framework. Compared to Neuroclear, Dual-Cycle includes two perpendicular views of the sample as input and uses prior knowledge on the estimated PSFs as a part of the degradation process within the framework. We have experimentally shown that Dual-Cycle achieves the state-of-the-art performance on synthetic and real data. While we only explored the dual-view setup in this work, our framework can be readily expanded into the multiple-view regime.

\section*{Acknowledgements}

This work was supported in part by the Czech Science Foundation grant GA21-03921S, the NSF
CAREER award CCF-2043134, the Fulbright commission under the Fulbright-Masaryk award, and by the Beckman Center for Advanced Light-Sheet Microscopy at Washington University in St.~Louis.

\bibliographystyle{IEEEtran}

\begin{thebibliography}{10}
\providecommand{\url}[1]{#1}
\csname url@samestyle\endcsname
\providecommand{\newblock}{\relax}
\providecommand{\bibinfo}[2]{#2}
\providecommand{\BIBentrySTDinterwordspacing}{\spaceskip=0pt\relax}
\providecommand{\BIBentryALTinterwordstretchfactor}{4}
\providecommand{\BIBentryALTinterwordspacing}{\spaceskip=\fontdimen2\font plus
\BIBentryALTinterwordstretchfactor\fontdimen3\font minus
  \fontdimen4\font\relax}
\providecommand{\BIBforeignlanguage}[2]{{%
\expandafter\ifx\csname l@#1\endcsname\relax
\typeout{** WARNING: IEEEtran.bst: No hyphenation pattern has been}%
\typeout{** loaded for the language `#1'. Using the pattern for}%
\typeout{** the default language instead.}%
\else
\language=\csname l@#1\endcsname
\fi
#2}}
\providecommand{\BIBdecl}{\relax}
\BIBdecl



\bibitem{stelzer2021light}
Ernst~HK Stelzer, Frederic Strobl, Bo-Jui Chang, Friedrich Preusser, Stephan
  Preibisch, Katie McDole, and Reto Fiolka,
\newblock ``Light sheet fluorescence microscopy,''
\newblock {\em Nature Reviews Methods Primers}, vol. 1, no. 1, pp. 1--25, 2021.

\bibitem{liu2022recovery}
Renhao Liu, Yu~Sun, Jiabei Zhu, Lei Tian, and Ulugbek~S Kamilov,
\newblock ``Recovery of continuous 3d refractive index maps from discrete
  intensity-only measurements using neural fields,''
\newblock {\em Nature Machine Intelligence}, pp. 1--11, 2022.

\bibitem{richardson1972bayesian}
William~Hadley Richardson,
\newblock ``Bayesian-based iterative method of image restoration,''
\newblock {\em JoSA}, vol. 62, no. 1, pp. 55--59, 1972.

\bibitem{lucy1974iterative}
Leon~B Lucy,
\newblock ``An iterative technique for the rectification of observed
  distributions,''
\newblock {\em The astronomical journal}, vol. 79, pp. 745, 1974.

\bibitem{wu2013spatially}
Yicong Wu, Peter Wawrzusin, Justin Senseney, Robert~S Fischer, Ryan
  Christensen, Anthony Santella, Andrew~G York, Peter~W Winter, Clare~M
  Waterman, Zhirong Bao, et~al.,
\newblock ``Spatially isotropic four-dimensional imaging with dual-view plane
  illumination microscopy,''
\newblock {\em Nature biotechnology}, vol. 31, no. 11, pp. 1032--1038, 2013.

\bibitem{kumar2014dual}
Abhishek Kumar, Yicong Wu, Ryan Christensen, Panagiotis Chandris, William
  Gandler, Evan McCreedy, Alexandra Bokinsky, Daniel~A Col{\'o}n-Ramos, Zhirong
  Bao, Matthew McAuliffe, et~al.,
\newblock ``Dual-view plane illumination microscopy for rapid and spatially
  isotropic imaging,''
\newblock {\em Nature protocols}, vol. 9, no. 11, pp. 2555--2573, 2014.

\bibitem{preibisch2014efficient}
Stephan Preibisch, Fernando Amat, Evangelia Stamataki, Mihail Sarov, Robert~H
  Singer, Eugene Myers, and Pavel Tomancak,
\newblock ``Efficient bayesian-based multiview deconvolution,''
\newblock {\em Nature methods}, vol. 11, no. 6, pp. 645--648, 2014.

\bibitem{temerinac2011multiview}
Maja Temerinac-Ott, Olaf Ronneberger, Peter Ochs, Wolfgang Driever, Thomas
  Brox, and Hans Burkhardt,
\newblock ``Multiview deblurring for 3-d images from light-sheet-based
  fluorescence microscopy,''
\newblock {\em IEEE Transactions on Image Processing}, vol. 21, no. 4, pp.
  1863--1873, 2011.

\bibitem{guo2020rapid}
Min Guo, Yue Li, Yijun Su, Talley Lambert, Damian~Dalle Nogare, Mark~W Moyle,
  Leighton~H Duncan, Richard Ikegami, Anthony Santella, Ivan Rey-Suarez,
  et~al.,
\newblock ``Rapid image deconvolution and multiview fusion for optical
  microscopy,''
\newblock {\em Nature biotechnology}, vol. 38, no. 11, pp. 1337--1346, 2020.

\bibitem{park2022deep}
Hyoungjun Park, Myeongsu Na, Bumju Kim, Soohyun Park, Ki~Hean Kim, Sunghoe
  Chang, and Jong~Chul Ye,
\newblock ``Deep learning enables reference-free isotropic super-resolution for
  volumetric fluorescence microscopy,''
\newblock {\em Nature Communications}, vol. 13, no. 1, pp. 1--12, 2022.

\bibitem{wu2021multiview}
Yicong Wu, Xiaofei Han, Yijun Su, Melissa Glidewell, Jonathan~S Daniels, Jiamin
  Liu, Titas Sengupta, Ivan Rey-Suarez, Robert Fischer, Akshay Patel, et~al.,
\newblock ``Multiview confocal super-resolution microscopy,''
\newblock {\em Nature}, vol. 600, no. 7888, pp. 279--284, 2021.

\bibitem{zhu2017unpaired}
Jun-Yan Zhu, Taesung Park, Phillip Isola, and Alexei~A Efros,
\newblock ``Unpaired image-to-image translation using cycle-consistent
  adversarial networks,''
\newblock in {\em Proceedings of the IEEE international conference on computer
  vision}, 2017, pp. 2223--2232.

\bibitem{huygens}
``Huygens psf distiller,'' \url{https://svi.nl/Huygens-PSF-Distiller},
\newblock Accessed: 2022-08-24.

\bibitem{de2003image}
Jacques~Boutet de~Monvel, Eric Scarfone, Sophie Le~Calvez, and Mats Ulfendahl,
\newblock ``Image-adaptive deconvolution for three-dimensional deep biological
  imaging,''
\newblock {\em Biophysical journal}, vol. 85, no. 6, pp. 3991--4001, 2003.

\bibitem{becker2019deconvolution}
Klaus Becker, Saiedeh Saghafi, Marko Pende, Inna Sabdyusheva-Litschauer,
  Christian~M Hahn, Massih Foroughipour, Nina J{\"a}hrling, and Hans-Ulrich
  Dodt,
\newblock ``Deconvolution of light sheet microscopy recordings,''
\newblock {\em Scientific reports}, vol. 9, no. 1, pp. 1--14, 2019.

\bibitem{isola2017image}
Phillip Isola, Jun-Yan Zhu, Tinghui Zhou, and Alexei~A Efros,
\newblock ``Image-to-image translation with conditional adversarial networks,''
\newblock in {\em Proceedings of the IEEE conference on computer vision and
  pattern recognition}, 2017, pp. 1125--1134.

\bibitem{huisken2004optical}
Jan Huisken, Jim Swoger, Filippo Del~Bene, Joachim Wittbrodt, and Ernst~HK
  Stelzer,
\newblock ``Optical sectioning deep inside live embryos by selective plane
  illumination microscopy,''
\newblock {\em Science}, vol. 305, no. 5686, pp. 1007--1009, 2004.

\bibitem{schindelin2012fiji}
Johannes Schindelin, Ignacio Arganda-Carreras, Erwin Frise, Verena Kaynig, Mark
  Longair, Tobias Pietzsch, Stephan Preibisch, Curtis Rueden, Stephan Saalfeld,
  Benjamin Schmid, et~al.,
\newblock ``Fiji: an open-source platform for biological-image analysis,''
\newblock {\em Nature methods}, vol. 9, no. 7, pp. 676--682, 2012.

\bibitem{fernandez2021fijiyama}
Romain Fernandez and C{\'e}dric Moisy,
\newblock ``Fijiyama: a registration tool for 3d multimodal time-lapse
  imaging,''
\newblock {\em Bioinformatics}, vol. 37, no. 10, pp. 1482--1484, 2021.

\end{thebibliography}


\end{document}